\documentclass[final,5p,times,twocolumn]{elsarticle}

\usepackage{amssymb}
\usepackage{amsthm}

\usepackage{hyperref}
\usepackage{siunitx} 
\usepackage{subfigure}
\usepackage{apalike}
\usepackage{url}
\usepackage{subfigure}
\usepackage{algorithm}
\usepackage{algpseudocode}

\journal{VISAPP 2020}

\begin{document}

\begin{frontmatter}



\title{Estimation of Muscle Fascicle Orientation in Ultrasonic Images}


\author[a,1]{Regina Pohle-Fr\"ohlich}
\ead{regina.pohle@hsnr.de}
\author[a,1]{Christoph Dalitz}
\ead{christoph.dalitz@hsnr.de}
\author[b,c]{Charlotte Richter}
\author[b,c]{Benjamin St\"{a}udle}
\author[b,c]{and Kirsten Albracht}

\address[a]{Institute for Pattern Recognition, Niederrhein University of Applied Sciences, Reinarzstr. 49, Krefeld, Germany}
\address[b]{Institute of Biomechanics and Orthopaedics, German Sport University Cologne,  Cologne, Germany}
\address[c]{Department of ﻿﻿Medical Engineering and Technomathematics, Aachen University of Applied Science, Germany}


\begin{abstract}
We compare four different algorithms for automatically estimating the muscle fascicle angle from ultrasonic images: the vesselness filter, the Radon transform, the projection profile method and the gray level cooccurence matrix (GLCM). The algorithm results are compared to ground truth data generated by three different experts on 425 image frames from two videos recorded during different types of motion. The best agreement with the ground truth data was achieved by a combination of pre-processing with a vesselness filter and measuring the angle with the projection profile method. The robustness of the estimation is increased by applying the algorithms to subregions with high gradients and performing a LOESS fit through these estimates.
\end{abstract}


\end{frontmatter}

\section{\uppercase{Introduction}}
\label{sec:introduction}

\begin{figure}[b!]
    \center\includegraphics[width=1.0\columnwidth]{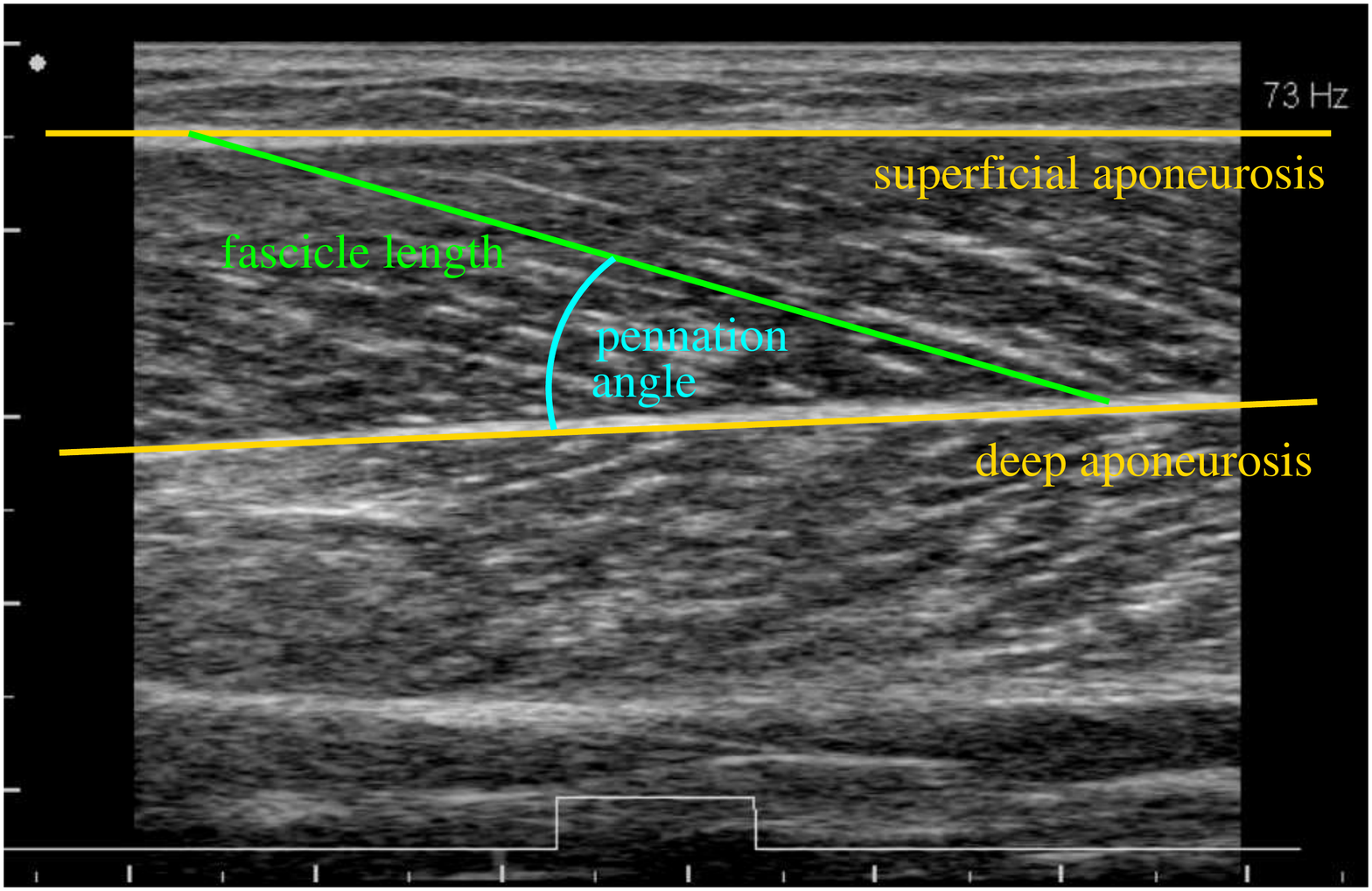}
    \caption{\label{fig:sampleimage} Annotated example of an ultrasonic image of the muscle gastrocnemius medialis.}
\end{figure}

Human movement results from a coordinated activation of the skeletal muscles. The muscle fascicle length and their change in length is critical for the force and efficiency of the muscle. It is thus necessary to measure fascicle length, which is usually done from ultrasonic images \cite{fukunaga97,fukunaga01,zhou12}. An example of a B-mode ultrasound image of the muscle gastrocnemius medialis recorded with an ALOKA Prosound $\alpha$7 can be seen in Fig.~\ref{fig:sampleimage}: the fascicles are spanned between the two aponeuroses.

As the fascicles are interrupted by noise and rarely are captured in their full length by the imaging process, their length must be computed from three different auxiliary observables: the position of the two aponeuroses and the fascicle orientation angle (pennation). Throughout the present paper, we make the simplifying assumption that both aponeuroses can be approximated by straight lines. The fascicle length can then be computed from the pennation angle at different positions on these lines. We thus only concentrate on the problem of finding the aponeuroses and estimating the pennation angle.

When the imaging is done while the muscle is in motion, the image quality can deteriorate due to variation of transducer skin contact and position \cite{aggeloussis10}. If the fascicle length or orientation estimation is not done manually, but semi-automatic or even fully automatic, this requires thus robust image processing methods for possibly noisy videos.

According to \cite{yuan20}, algorithms for automatically estimating the fascicle orientation can be divided  into three different approaches. In the first category, the tracking is semi-automatic by following a manually marked indicidual fibers in subsequent frames. This can be done, for example by calculating the optical flow, like in the UltraTrack software \cite{farris16}. A disadvantage of these methods is the cumulative error, which requires manual correction after several frames. In addition, misalignments may result due to significant changes in the appearance and intensity of the structures between successive frames. These problems occur particularly with large displacement fields due to fast motion and insufficient sampling rates of most currently available commercial devices.

Methods based on texture feature detection form the second category, which includes Hough transform \cite{zhou08}, Radon transform \cite{zhao11}, or vesselness filter \cite{rana09}. The disadvantage of these methods is that the result of the angle estimation may be distorted by speckle noise and intramuscular blood vessels, which modify the characteristics of the muscle fascicles.

The third category includes deep learning approaches. Cunningham proposed deep residual \cite{cunningham17} and convolution neural networks \cite{cunningham18} to estimate the muscle fascicle orientation. One problem with using deep learning methods is that they require a large amount of manually measured image data to achieve good results. Another difficulty is the dependence of the image acquisition and the image distortions on the ultrasound transducer, so that adjusted data sets are required.

In the present article, we compare two established methods from the literature with two new approaches to determine the orientation of textures. As established methods, we consider vesselness filtering \cite{rana09} and Radon transform \cite{zhao11}.  We compare these with the very recently proposed gray value cooccurence matrix based texture orientation estimation \cite{zheng18} and the calculation of the angle using the projection profile \cite{dalitz08}. The latter method has been used for some time in document image analysis for estimating the rotation of binary documents. Here we demonstrate that it can be used for gray level images, too.

In order to evaluate the quality of the different algorithms, we have compared their results with manual estimations of the pennation angle by different expert observers. As evaluation criteria, we utilized the intra-class correlation and the mean absolute percentage with respect to the inter-observer average, and the percentage of results within the inter-observer range.

This article is organized as follows: in section \ref{sec:roi} \& \ref{sec:direction} we describe the implemented algorithms, section \ref{sec:evaluation} describes the evaluation method, section \ref{sec:results} discusses the results and compares the algorithm performances, and in section \ref{sec:conclusion} we draw some conclusions and give recommendations for a practical utilization of the algorithms.

\section{\uppercase{Region of Interest Extraction}}
\label{sec:roi}

To determine the region of interest (ROI), each video frame is evaluated separately. Firstly, the two black areas (see Fig.\ref{fig:sampleimage}) are removed. Then, for a reinforcement of the aponeuroses a vesselness filtering (see section \ref{sec:direction:vesselness}) is carried out. Then, Otsu's thresholding method is used generate a binary image of the filtered image. In the result, the two largest segments which correspond to the two aponeuroses are selected. Straight lines are fitted to the lower segment border of the superficial aponeurosis and to the upper segment border of the deep aponeurosis using the least squares method. The height of the ROI resulted from the difference between the smallest $y$-value of the lower aponeurosis minus 10 pixels and the largest $y$-value of the upper aponeurosis plus 10 pixels. The width of the ROI is calculated from the width of the image minus a safety area of 10 pixels to the left and right borders. This ensures that the ROI is always positioned within the muscle. As the noise level or the orientation angle may vary over the entire ROI, we additionally subdivided the entire region horizontally into eight overlapping subregions. For a fully automated process, it would be necessary to automatically pick the subregion with the ``best'' image quality. To characterize this quality, we have computed, for every subregion, the gray value variance as a measure for contrast, the mean gradient value and the maximum value of the histogram of the gradients as measures for edge sharpness.

\section{\uppercase{Fascicle Direction Estimation}}
\label{sec:direction}

For the determination of the fiber orientation we used different methods, which are described in the following. These methods were either applied directly to the ROI or a pre-processing step was used for fascicle enhancement. For pre-processing, a Vesselness filter or Radon transformation was optionally applied for image enhancement. Tbl.~\ref{tbl:combinations} shows the investigated combinations for pre-processing and fascicle orientation estimation. 
\begin{table}[t]
\centering\begin{tabular}{c|ccc} 
orientation & \multicolumn{3}{c}{pre-processing} \\
estimation & none & Frangi & Radon \\
\hline  
Frangi & - & x & -\\
Radon & - & - & x\\
GLCM & x & x & x\\
projections & x & x &x\\
\end{tabular}
\caption{Tested combinations for pre-processing and fascicle orientation estimation. ``Frangi'' denotes the vesselness filter.}
\label{tbl:combinations}

\end{table}

\subsection{Radon Transform}
\label{sec:direction:radon}

\begin{figure}[t]
  \subfigure[\label{fig:preprocessing:raw}raw data]{\includegraphics[width=1.0\columnwidth]{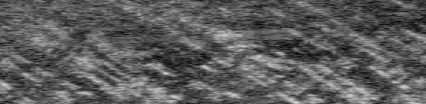}}
 \subfigure[\label{fig:preprocessing:frangi}vesselnes filter (Frangi)]{\includegraphics[width=1.0\columnwidth]{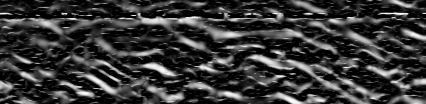}}
 \subfigure[\label{fig:preprocessing:radon}Radon transform]{\includegraphics[width=1.0\columnwidth]{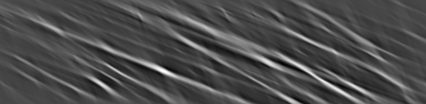}}
    \caption{\label{fig:preprocessing} Effect of filtering with the vesselness filter or the Radon transform on an image recorded during running movement.}
\end{figure}

The Radon transformation determines the line integral of the function $f(x,y)$ along all straight lines of the $xy$ plane. For each of these straight lines one can consider the Radon transform $R$ as a projection of the function $f(x,y)$ onto a straight line perpendicular to it. For this reason it was used by \cite{zhao11}, \cite{yuan20} to determine the orientation of the muscle fibers in ultrasound images. It should be noted that the radon transformation cannot only be used for direct angle estimation, but also merely as a pre-processing operation to reinforce fascicles. Such a pre-processed image $E$ with an enhancement of the linear structures in the initial image $I$ is achieved by applying the following equation:
\begin{equation}
E=R^{-1}(\mbox{sign}(R(I))\cdot R(I)^2)
\end{equation} 
where $R$ is the Radon transform and $R^{-1}$  is the inverse Radon transform. The result of the Radon transform based enhancement is shown in Fig.~\ref{fig:preprocessing:radon}. The angle of the fascicle orientation resulted from the position of the maximum of the radon transformed. In our tests, we calculated the radon transformation only for an angular range of 15 to 70 degrees in which the actual values vary to exclude errors due to the orientation of the speckle pattern.

\subsection{Vesselness Filter}
\label{sec:direction:vesselness}

Muscle fascicles appear in ultrasound images as vessel-like tubular structures, so that in \cite{rana09} the multiscale vesselness filter developed by Frangi  \cite{frangi98} was used to enhance them.

In the first step of this filter, images are convolved with Gaussian kernels. Then the Hessian matrix of these convolved images is computed. Their eigenvalues provide information related to the direction of line-like structures. The eigenvector in the direction of the smallest eigenvalue yields the orientation angle at the respective pixel position. For our tests we used the implementation in {\em libfrangi}\footnote{\url{https://github.com/ntnu-bioopt/libfrangi}} whereby we only allowed angles within our chosen range of 15 to 70 degrees in order to suppress responses from dominating horizontal or vertical structures. All values outside this range were set to zero in the result image. To estimate a total orientation angle from all the local angles estimated at non-zero pixels, we estimated the angle distribution with a kernel density estimator with ``Silverman's rule of thumb'' \cite{sheather04} and determined the angle maximizing this density.

Like the radon transform, the vesselness filter can alternatively also merely be used as a pre-processing operation for enhancing fascicle structures. An example is shown in Fig.~\ref{fig:preprocessing:frangi}.  

\subsection{Projection Profile}
\label{sec:direction:projection}
The projection profile method \cite{dalitz08} estimates the orientation angle $\alpha$ as the angle with the highest variation of the skewed projection profile
\begin{equation}
h_{\alpha}(y)=\sum_{x=-\infty}^{x=\infty} f(x \cos \alpha - y \sin \alpha, x \sin \alpha + y \cos \alpha)
\end{equation} 
where $f(x,y)$ is the gray value of the ultrasound image at position $(round(x), round(y))$ and zero outside the image. The variation of this profile is defined as
\begin{equation}
V(\alpha)=\sum_{y=-\infty}^{y=\infty} [h_{\alpha}(y+1) - h_{\alpha}(y) ]^2
\end{equation} 
In our implementation we calculate the variation for an angle range from 15 to 70 degrees with a step width of 0.5 degrees, which corresponds to the possible angles occurring for our recording conditions. Then we select the angle corresponding to the highest variation.

\subsection{Graylevel Cooccurence}
\label{sec:direction:cooccurrence}
The gray level cooccurence matrix (GLCM) represents an estimate of the probability that a pixel at position $(x, y)$ in an image  with a graylevel $g_1$ has a graylevel $g_2$ at position $(x+dx, y+dx)$. The GLCM has a size of $g_{max} \times g_{max}$, whereby $g_{max}-1$ is the maximum of the gray levels in the image. If arbitrary relative positions are used to calculate the GLCM, the texture orientation can be estimated. Zheng \cite{zheng18} applied this method to evaluate SAR images of the sea surface. For the calculation of the GLCM, we utilized in the method that Zheng et al.~called ``scheme 1''. If the shift vector $(dx,dy)=(r\cdot\cos\alpha, r\cdot\sin\alpha)$ corresponds with the texture orientation, the diagonal elements of the GLCM attain high values. For the estimation of the fascicle orientation, we apply the criterion suggested in \cite{zheng18}, i.e., the degree of concentration  $C$ of larger elements of the GLCM with respect to the diagonal line:
\begin{equation}
C(r,\alpha)=\sum_{m=0}^{g_{max}-1} \sum_{n=0}^{g_{max}-1} (m-n)^2 \cdot GLCM(m,n;r, \alpha)
\end{equation} 
The weight $(m-n)^2$, which increases with increasing distance of the matrix element from the diagonal, results in smaller values for images with a strong line structure if the angle $\alpha$ corresponds to the orientation of this structure. In our experiments we used a maximum $r$ of 40 and an angle range of 15 to 70 degrees. The used angle corresponded to the angle $\alpha$ with the lowest concentration value.

\subsection{Local Regression}
\label{sec:direction:regression}

Due to the noisy nature of the images, the angle estimate can fluctuate considerably between adjacent frames and subregions. It is thus natural to seek a more robust angle estimate by means of local regression. To this end, we optionally apply Cleveland \& Devin's LOESS method \cite{cleveland88}, which is a distance weighted least squares fit over the $k$ nearest neighbors with weight
\begin{equation}
  \label{eq:loess}
  W_h(z)=\left\{\begin{array}{ll}\left(1-\left(z/h\right)^3\right)^3 & \mbox{ for }|z|<h\\
  0 & \mbox{ otherwise}\end{array}\right.
\end{equation}
where $h$ is the distance to the $k$-th nearest neighbor. In our case, the predictor is the frame number and the dependent variable is the pennation angle.

\section{\uppercase{Evaluation Method}}
\label{sec:evaluation}

\noindent In order to evaluate and compare the different algorithms, we have asked three different experts to manually draw the aponeurosis and fascicle orientation into ultrasonic images with the user interface of the UltraTrack software \cite{cronin11}. The images were taken from two different videos, which were recorded each with an ALOKA Prosound $\alpha$7 for five consecutive stance phases (touchdown to toe-off) of the left foot during walking (video ``W'') and running (video ``R''). This resulted in a total of 425 different frames. The muscle fascicles in the R video were less clearly visible tan in the W video due to the shakier transducer skin contact during running.

\begin{figure}[t]
    \center\includegraphics[width=1.0\columnwidth]{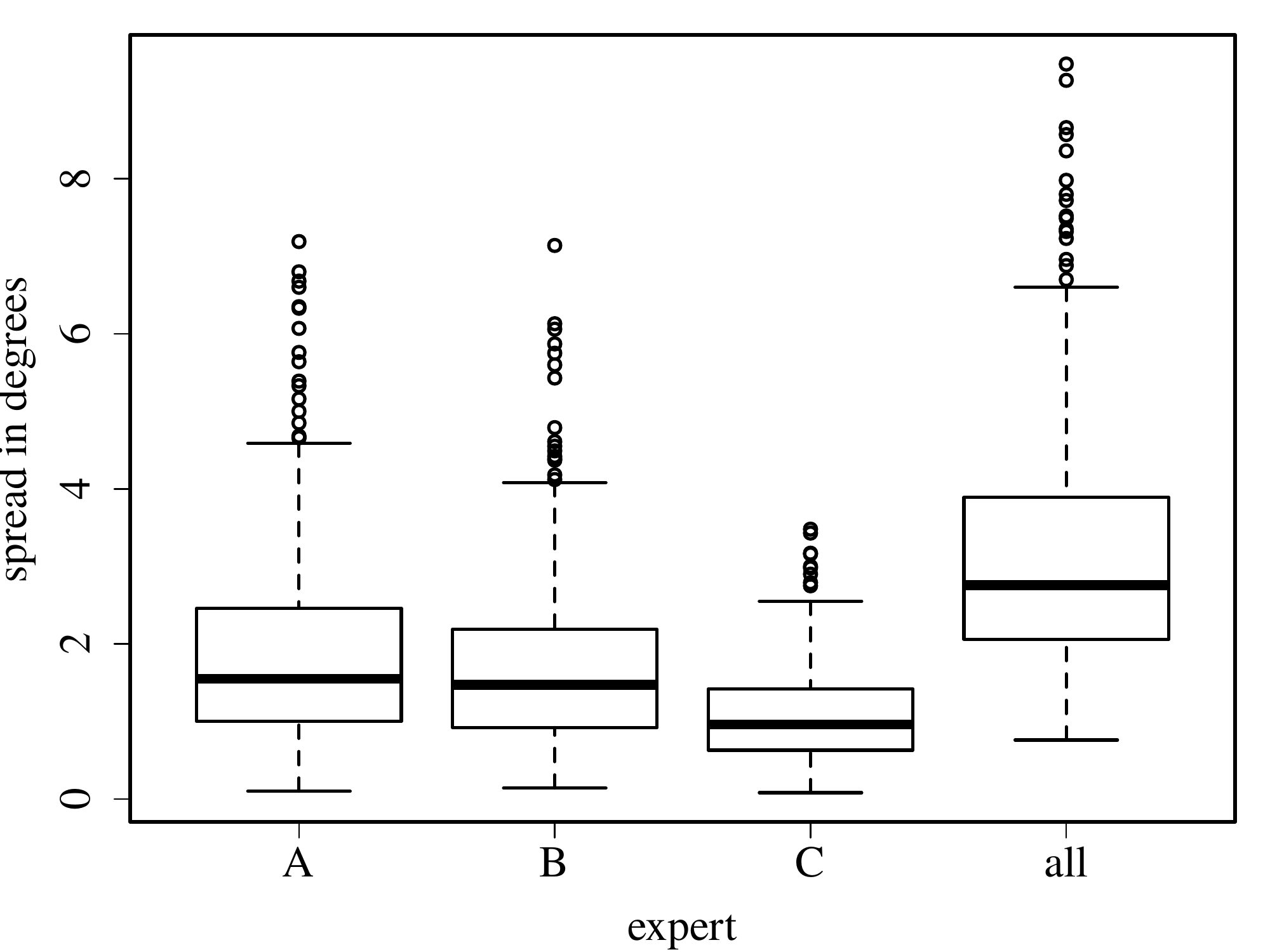}
    \caption{\label{fig:manual-angle-spread} Spread per frame of the pennation angle estimates of the three experts.}
\end{figure}

Each frame was examined three times by every expert, but on different days. We thus had nine different manually estimated angles for each frame. This was done to estimate the accuracy of the expert opinion. The intra-class correlation ICC3 \cite{shrout79} between the experts' angle estimations was 0.97, which means that there was good agreement among the experts which angles were higher and which were lower. On the other hand, the average angle spread per frame was $1.9^\circ$ for expert $A$, $1.7^\circ$ for expert $B$, $1.1^\circ$ for expert C, and $3.2^\circ$ over all experts. Box-Plots for the spread distribution can be seen in Fig.~\ref{fig:manual-angle-spread}. The spread between experts was thus considerably greater than within each expert, and we conclude that we cannot expect an algorithm to estimate the angle with an accuracy greater than about two degrees.

Part of the inter- and intra-observer variation can be explained by varying fascicle orientations for different image regions. We therefore split the region of interest into eight slightly overlapping subregions and ran the algorithms on each subregion plus on the entire region. For each algorithm, we then measured the following performance indicators for each of these nine regions:
\begin{itemize}
\item the intra-class correlation (ICC3) with the inter-observer average; this measures how well the estimated angles follow the curve shape
\item the mean absolute error (MAE) with respect to the inter-observer average; this measures the overall error in the estimation in degrees
\item the percentage of values inside the inter-observer range (hit)
\end{itemize}

\section{\uppercase{Results}}
\label{sec:results}
As the pennation angle is defined as the angle between the deep aponeurosis and the muscle fascicles, there are two possible sources of error for its estimation: errors in the estimation of the aponeurosis' slope, and in the  estimation of the fascicle orientation. We therefore first evaluated the aponeurosis estimation, and then the estimation of the pennation angle. Moreover, to derive recommendations for pre-processing filtering, we report results for the different combinations of pre-processing and estimation algorithms listed above in Tbl.~\ref{tbl:combinations}.

\subsection{Aponeurosis slope}
In video ``R'', the deep aponeurosis was very close to a straight line, and algorithm and expert opinion about its slope angle was in good agreement: ICC3=$0.926$, MAE=$0.286^\circ$, hit=$71\%$.

\begin{figure}[b!]
  \centering\includegraphics[width=\columnwidth]{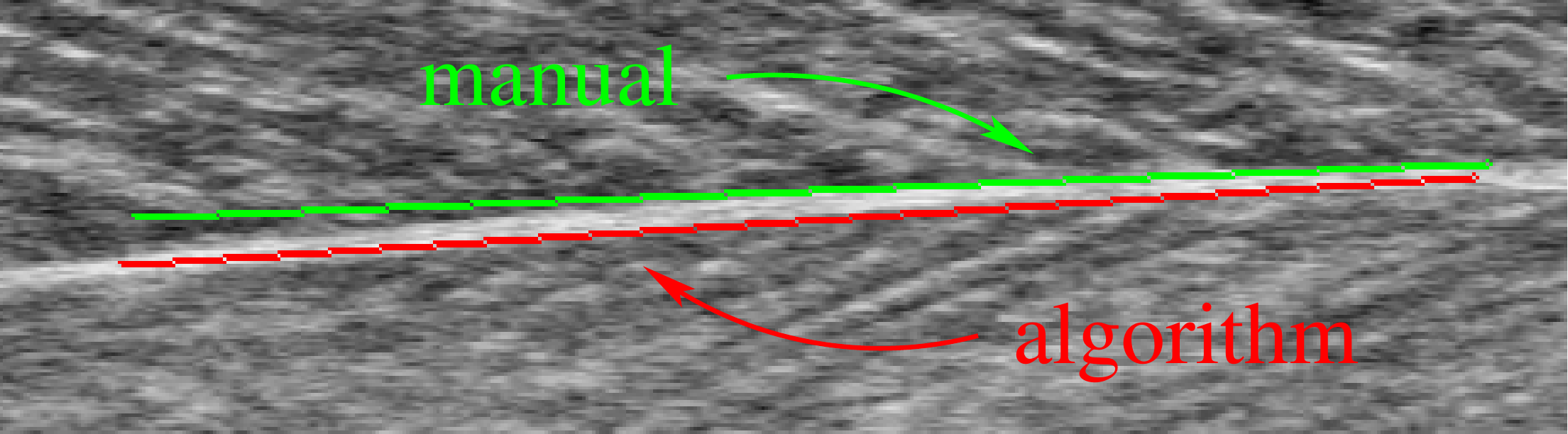}
  \caption{\label{fig:curved-apo}The aponeurosis curvature in video ``W'' leads to a small difference in the aponeurosis slope estimation.}
\end{figure}

In video ``W'', the deep aponeurosis was curved slightly (see Fig.~\ref{fig:curved-apo}) and the experts tended to estimate the slope at the right end, whilst the algorithm computed an average slope over its entire width. This had the effect that the automatic estimate of its slope angle was on average one degree greater than the expert opinion: ICC3=$0.764$, MAE=$1.081^\circ$, hit=$4\%$.

As the decision at which position the tangential angle of the aponeurosis is measured is somewhat arbitrary, we conclude that the aponeurosis slope angle is estimated by our algorithm within the possible accuracy. The difference in slope estimation has no effect for video ``R'', but for video ``W'' it leads to a systematic difference of about one degree for the pennation angle, i.e., the automatically estimated pennation angle in video ``W'' should be about one degree greater than the manually estimated angle.

\begin{figure*}[t]
  \centering\subfigure[\label{fig:eineregion:W}walking (video W)]{\includegraphics[width=0.4\textwidth]{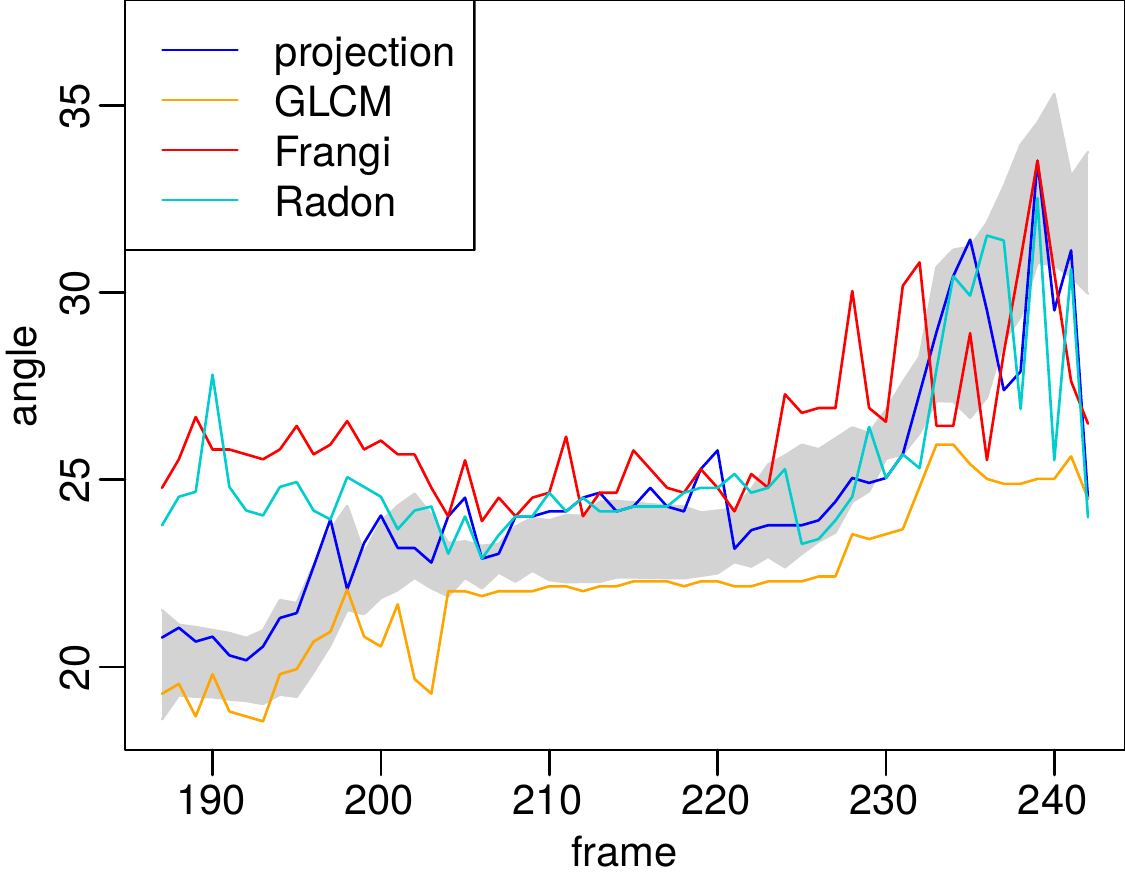}}\hspace{1em}
  \subfigure[\label{fig:eineregion:R}running (video R)]{\includegraphics[width=0.4\textwidth]{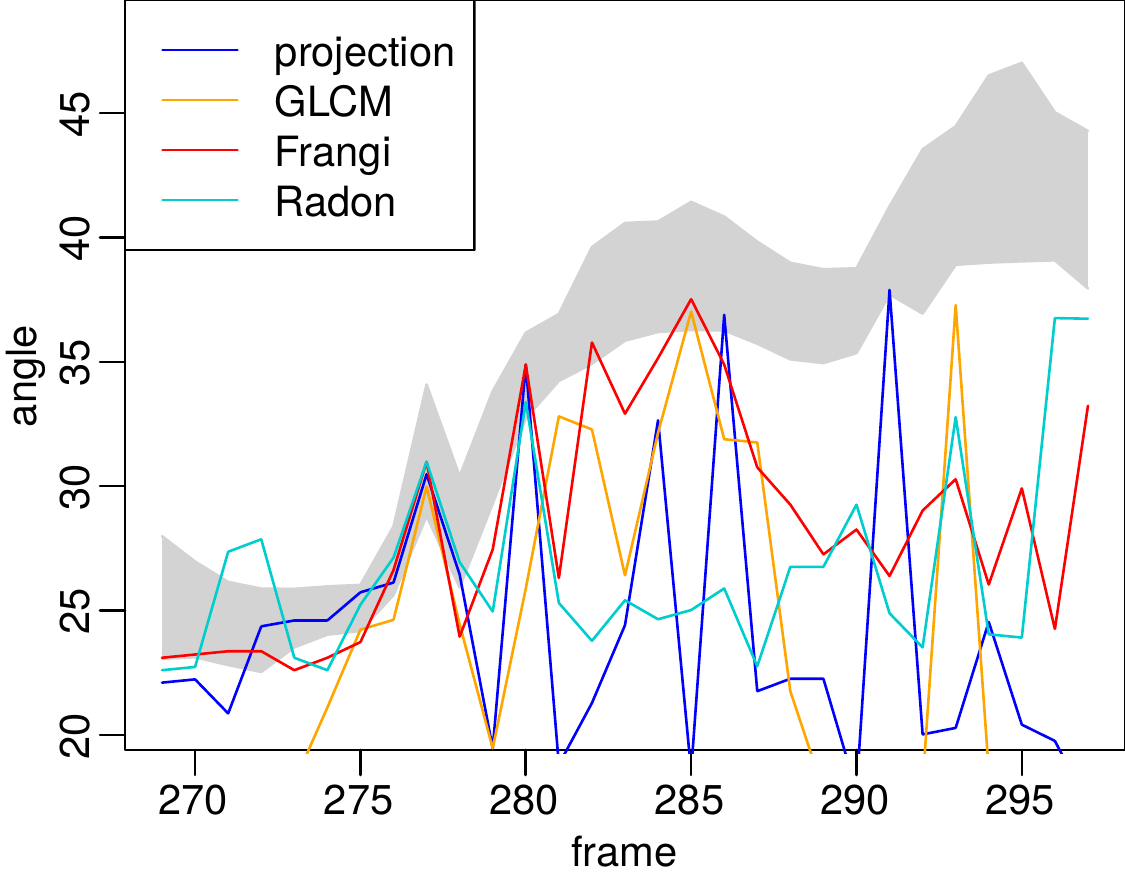}}
  \caption{\label{fig:eineregion}Angle estimations of the different algorithms applied to the {\em entire region} of interest for two typical steps of motion. The gray area is the inter-observer range.}
\end{figure*}

\begin{figure*}[t]
  \centering\subfigure[\label{fig:besteloess:W}walking (video W)]{\includegraphics[width=0.4\textwidth]{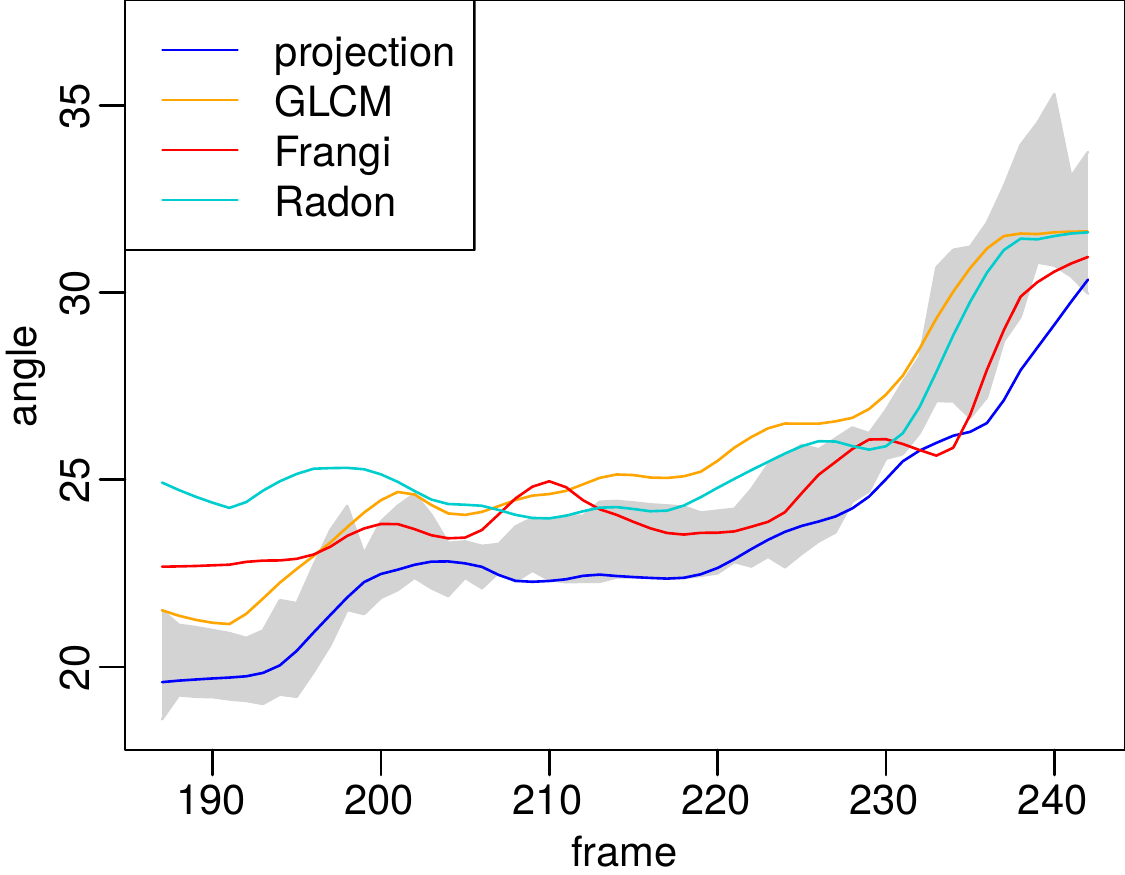}}\hspace{1em}
  \subfigure[\label{fig:besteloess:R}running (video R)]{\includegraphics[width=0.4\textwidth]{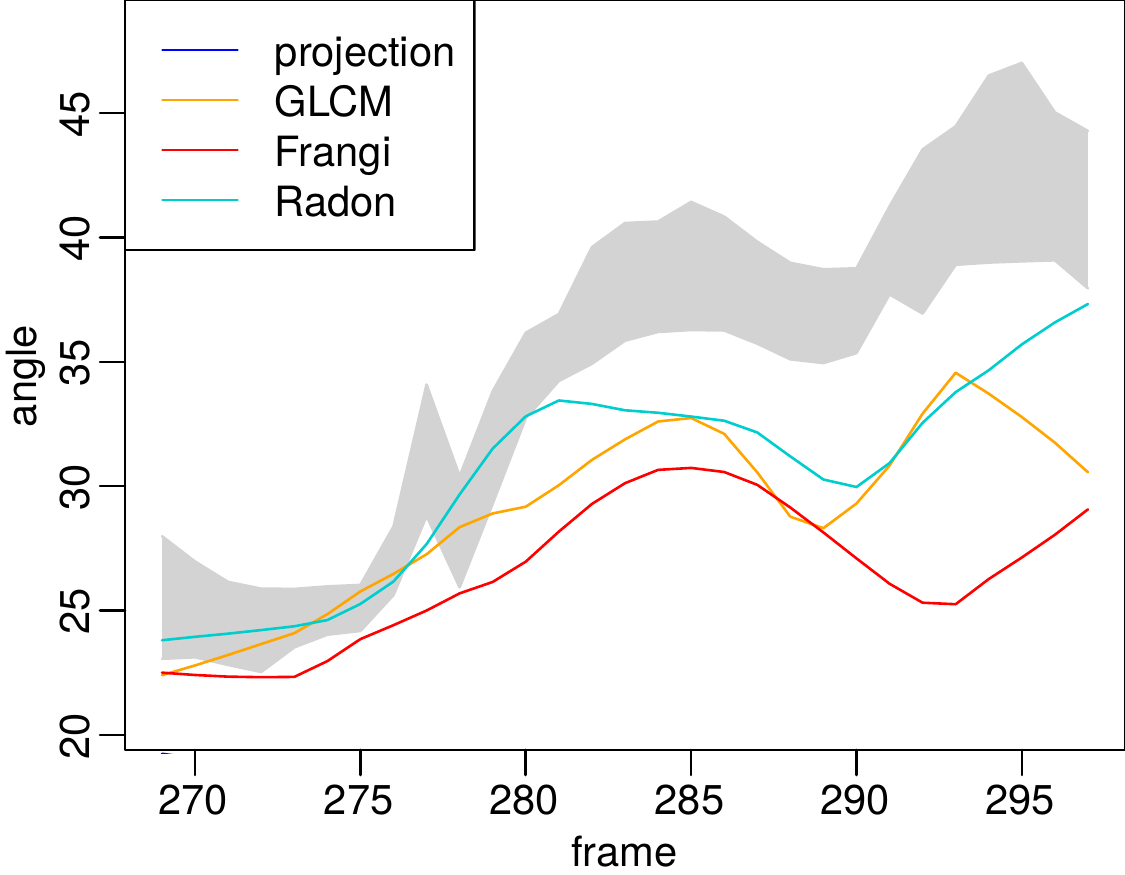}}
  \caption{\label{fig:besteloess}Angle estimations of the different algorithms applied to the three {\em regions with the highest mean gradient} and with LOESS fitting for two typical steps of motion. For video R, the projection method was so far off that its values do not fall into the displayed angle range. The gray area is the inter-observer range.}
\end{figure*}

\subsection{Pennation angle}
For the pennation angle, we have evaluated two different approaches to its estimation. The first approach models the angle as a single texture feature over the entire ROI, whilst the second approach models it as locally and statistically varying and applies a LOESS fit over subregions of neighboring frames.

\subsubsection{Entire region}
It turned out that the results were very different for the two videos: for all algorithms, all performance indices were considerably better on the less noisy video ``W'' (see Tbl.~\ref{tbl:eineregion}). The best performing algorithm was the projection profile method, followed by the GLCM. As can be seen in Fig.~\ref{fig:eineregion:W}, the angles estimated by the other two algorithm follow the curve shape with lesser agreement, which corresponds to poorer ICC3 values in Tbl.~\ref{tbl:eineregion}.

\begin{table}[t]
 \centering\begin{tabular}[b]{l|c|r|r|r}
   {\em algorithm} & {\em video} & {\em ICC3} & {\em MAE} & {\em hit}\\\hline
   projection & W & 0.871 & 1.231$^\circ$ & 58\% \\
   & R & -0.003 & 9.335$^\circ$ & 35\% \\\hline
   GLCM & W & 0.784 & 2.221$^\circ$ & 33\% \\
   & R & 0.180 & 10.437$^\circ$ & 10\% \\\hline
   Frangi & W & 0.552 & 2.998$^\circ$ & 17\% \\
   & R & 0.524 & 5.493$^\circ$ & 33\% \\\hline
   Radon & W & 0.540 & 2.309$^\circ$ & 42\% \\
   & R & 0.430 & 6.506$^\circ$ & 32\%
 \end{tabular}
 \caption{\label{tbl:eineregion}Angle estimation performance indices of the different algorithms applied to the {\em entire region} of interest.}
\end{table}

For video ``R'', however, neither of the algorithms yielded satisfying results, as can be concluded from the poor performance indices in Tbl.~\ref{tbl:eineregion} and the random fluctuations of the estimated angles inf Fig.~\ref{fig:eineregion:R}.

\subsubsection{LOESS fit over subregions}
To obtain a more robust angle estimator, we calculated the estimates for eight subregions, selected the ``best'' three subregions per frame and made a LOESS fit over these subregions including the eight neighboring frames. As our predictor was the frame number, the distance $z$ in Eq.~(\ref{eq:loess}) was measured in frame numbers and the number of neighbors was $k=27$.

This raises the question, how the ``best'' subregions are selected for each frame. A human expert would focus on a region in which the fascicles are clearly visible, i.e.~a region with high contrast or sharp edges. The three criteria listed in section \ref{sec:roi} try to measure this property. It turned out that the actual criterion has a smaller effect than the choice of algorithm. We thus present the results that use the highest mean gradient as a criterion for the ``best'' subregions; the results for the other criteria are similar.

\begin{figure*}[t]
  \centering\subfigure[\label{fig:preproc:W}walking (video W)]{\includegraphics[width=0.4\textwidth]{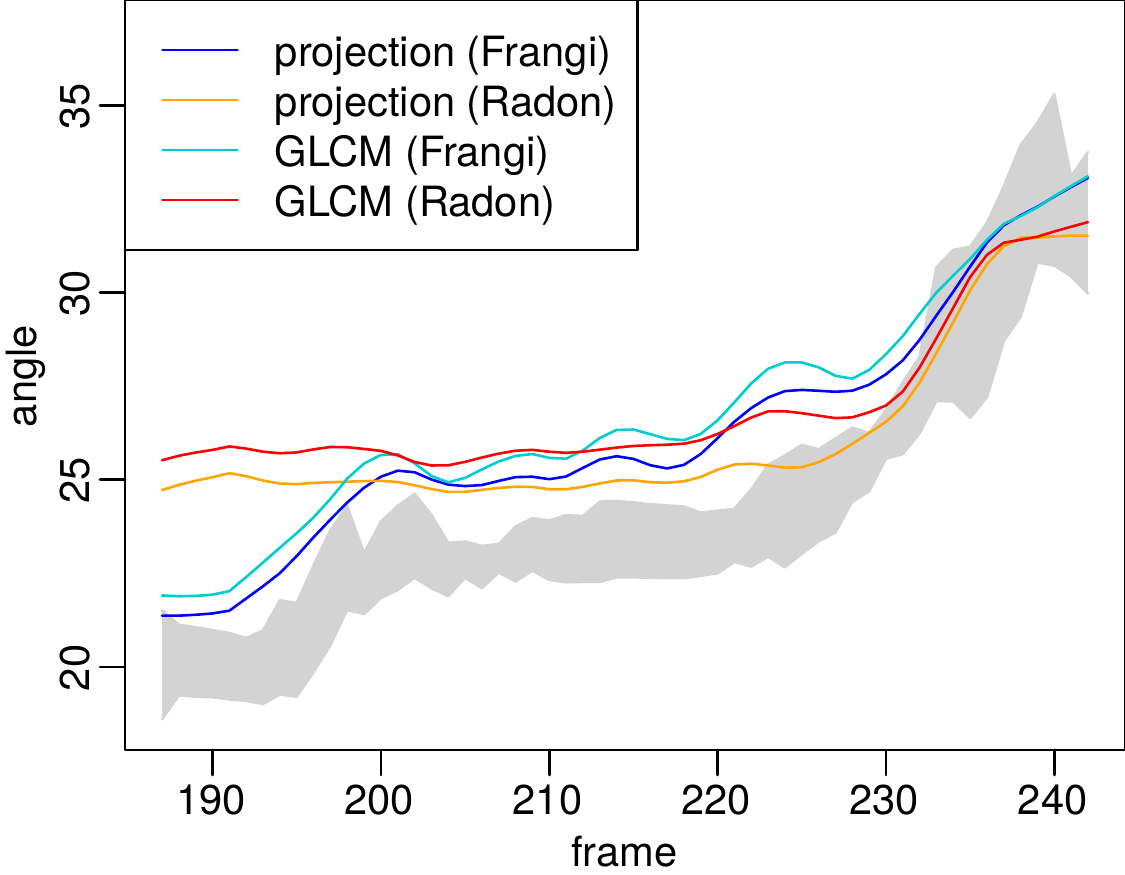}}\hspace{1em}
  \subfigure[\label{fig:preproc:R}running (video R)]{\includegraphics[width=0.4\textwidth]{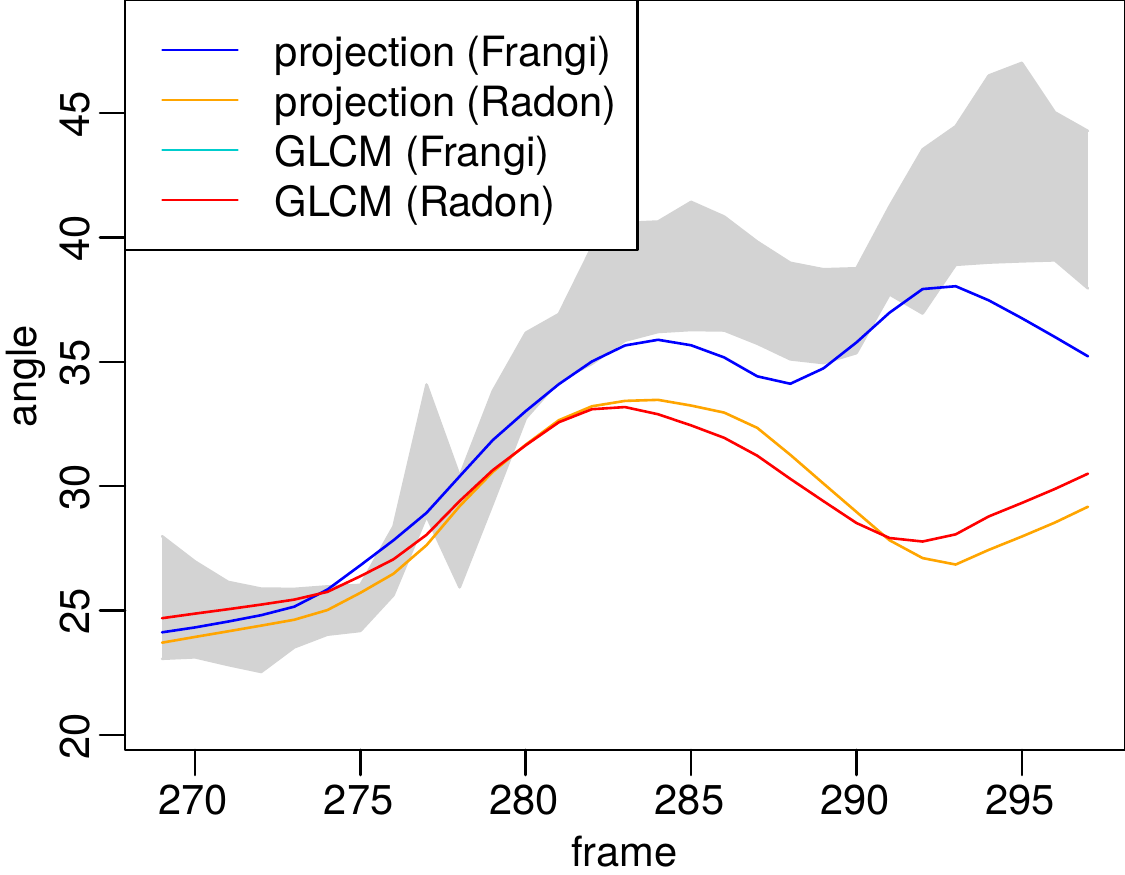}}
  \caption{\label{fig:preproc}Effect of the pre-processing filters (in parentheses) on angle estimations applied to the three {\em regions with the highest mean gradient} and with LOESS fitting for two typical steps of motion. For video R, the GLCM method with Frangi (vesselness filter) pre-processing was so far off that its values do not fall into the displayed angle range. The gray area is the inter-observer range.}
\end{figure*}

\begin{table}[t]
 \centering\begin{tabular}[b]{l|c|r|r|r}
   {\em algorithm} & {\em video} & {\em ICC3} & {\em MAE} & {\em hit}\\\hline
   projection & W & 0.975 & 0.548$^\circ$ & 86\% \\
   & R & 0.096 & 12.895$^\circ$ & 1\% \\\hline
   GLCM & W & 0.926 & 1.567$^\circ$ & 27\% \\
   & R & 0.848 & 4.122$^\circ$ & 29\% \\\hline
   Frangi & W & 0.733 & 1.704$^\circ$ & 49\% \\
   & R & 0.635 & 6.688$^\circ$ & 14\% \\\hline
   Radon & W & 0.804 & 2.007$^\circ$ & 41\% \\
   & R & 0.868 & 3.357$^\circ$ & 36\% \\
 \end{tabular}
 \caption{\label{tbl:besteloess}Angle estimations of the different algorithms applied to the three {\em regions with the highest mean gradient} and with LOESS fitting.}
\end{table}

As can be seen from Tbl.~\ref{tbl:besteloess}, the LOESS fit improves the performance indices in almost all cases. One notable exception is the projection profile method for video ``R'': in this case the angle estimates were so far off that they even fell outside the range of Fig.~\ref{fig:besteloess:R}, although this algorithm performed best on video ``W''. We thus conclude that the projection profile method should be used in combination with a pre-processing filter because it is not robust with respect to high levels of noise.

\subsection{Effect of pre-processing}

\begin{table}[t]
 \centering\begin{tabular}[b]{l|c|r|r|r}
   {\em algorithm} & {\em video} & {\em ICC3} & {\em MAE} & {\em hit}\\\hline
   projection & W & 0.946 & 1.962$^\circ$ & 25\% \\
   (with Frangi) & R & 0.946 & 1.871$^\circ$ & 62\% \\\hline
   projection & W & 0.755 & 2.231$^\circ$ & 31\% \\
   (with Radon) & R & 0.665 & 4.500$^\circ$ & 29\% \\\hline
   GLCM & W & 0.914 & 2.408$^\circ$ & 21\% \\
   (with Frangi) & R & 0.058 & 39.699$^\circ$ & 0\% \\\hline
   GLCM & W & 0.718 & 2.912$^\circ$ & 20\% \\
   (with Radon) & R & 0.695 & 4.316$^\circ$ & 23\% \\
 \end{tabular}
 \caption{\label{tbl:preproc}Angle estimations after pre-processing applied to the three {\em regions with the highest mean gradient} and with LOESS fitting.}
\end{table}

To see whether using the Radon transform or the vesselness filter (``Frangi'') as a pre-processing operation improves the performance of the other algorithms, we have first applied these filters and then utilized the same LOESS approach as in the preceding subsection. As can be seen from Tbl.~\ref{tbl:preproc}, this did not improve the performance of the GLCM with respect to Tbl.~\ref{tbl:besteloess}, but for the projection profile method, pre-processing with a vesselness filter seriously improved the results for video ``R''. Overall, the combination ``vesselness filter and projection profile method'' was the best performing algorithm, followed secondly by the GLCM without pre-processing.

\section{\uppercase{Conclusions}}
\label{sec:conclusion}
Based upon our experimental evaluation, we recommend two possible algorithms for estimating the pennation angle in ultrasonic images of muscles. The best performing algorithm was a combination of the vesselness filter as a pre-processing operation with the projection profile method for angle estimation. This algorithm achieved an intra-class correlation close to one and had a mean average error less than two degrees. The second best algorithm was based on the gray level cooccurance matrix (GLCM).

Both the robustness and accuracy of the angle estimates are considerably improved by a LOESS fit over neighboring frames and the subregions with the best visible edges. In our study, we have selected these regions automatically on basis of the mean absolute value of the gradient within the subregion.

In practice, if a semi-automatic processing is possible, the region selection process could alternatively done by an expert user. This would also have the benefit that the fascicle {\em length} computation can be based on the selected region. This is of relevance, because the fascicle length is not well defined if the superficial and the deep aponeuroses are not parallel. In this case, a hint by an expert user is necessary in any case where to set an anchor point of the line used for computing the fascicle length, which could be chosen, e.g., as the mid point of the user selected region.

\section*{\uppercase{Acknowledgments}}

\noindent Parts of this study were financially supported by the German Federal Ministry of Economic Affairs and Energy under grant no. 50WB1728.

\bibliographystyle{elsarticle-num}
\bibliography{fiber-orientation}

\end{document}